# Scalar Quantization for Audio Data Coding

Boris D. Kudryashov, Anton V. Porov, and Eunmi L. Oh

*Abstract*— This paper is concerned with scalar quantization of transform coefficients in an audio codec. The generalized Gaussian distribution (GGD) is used as an approximation of one-dimensional probability density function for transform coefficients obtained by modulated lapped transform (MLT) or modified cosine transform (MDCT) filterbank. The rationale of the model is provided in comparison with theoretically achievable rate-distortion function. The rate-distortion function computed for the random sequence obtained from a real sequence of samples from a large database is compared with that computed for random sequence obtained by a GGD random generator. A simple algorithm of constructing the Extended Zero Zone (EZZ) quantizer is proposed. Simulation results show that the EZZ quantizer yields a negligible loss in terms of coding efficiency compared to optimal scalar quantizers. Furthermore, we describe an adaptive version of the EZZ quantizer which works efficiently with low bitrate requirements for transmitting side information.

## I. INTRODUCTION

Most of perceptual audio codecs are based on the cosine-modulated filterbanks. Previous researches show that high transform gains can be obtained from such filterbanks with reasonable implementation complexity [1]–[3]. Typically, psychoacoustic module in perceptual audio codec is used to optimize the allocation of bit resource across subbands of an audio signal. Psychoacoustic module takes into account human ear sensitivity to detect distortions of an original sound, depending on frequency range, amplitude, neighboring (in frequency or time domain) sounds, and so on.

Critically sampled filterbank for each frame of $N$ time-domain samples outputs $N$ frequency-domain spectrum coefficients that are to be quantized and lossless encoded.

The quantization module is one of the most important modules of the audio codec. The quantizer receives the transform coefficients from the filterbank and the required quantization precision from the PAM module (see Fig.1).

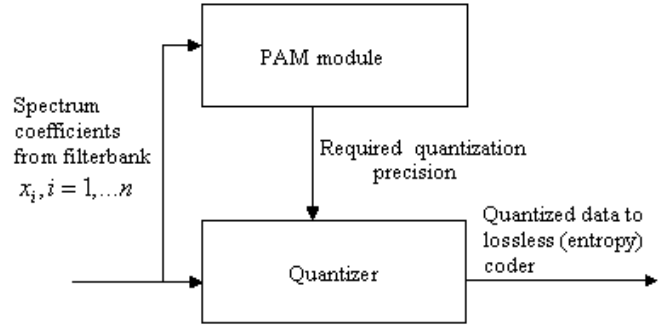

Fig. 1. Quantization module for audio.

The output of the quantizer is further processed by a lossless entropy encoder. For example, properly designed arithmetic encoder, as one of entropy encoders, can provide the same coding rate as entropy of the quantized data. Therefore, the entropy of quantized data can be considered as criteria for quantization analysis. The goal of this paper is to develop a quantizer that produces output data with a minimum entropy value under a given restriction on the quantization precision.

In general, there are two competing classes of quantizers: scalar and vector quantizers [4], [5]. Within each class we can also choose among different types of quantizers. The optimal choice, of course, depends on the range of bit rates and on the source model. However, there are some important practical limitations which make scalar quantization more favorable. The first is coding and reconstruction complexity. The coding gain of vector quantization over scalar quantization grows slowly with the quantization dimension, whereas memory consumption and computational complexity grows exponentially [4], [5].

One more argument in favor of scalar quantization is related to the adaptation of a quantizer to dynamic changes in source data probability distribution. Typically, the codebook of vector quantization is constructed from some training data set of audio signals. This codebook might be inefficient for audio signals, because we can have a very unusual input that is quite different from data set used in the construction of the codebook. This might not happen for scalar quantizer since each component is processed separately.

Yet another (and probably the most important one for low-rate audio codecs) argument is more sophisticated. It takes into account specific properties of the generalized Gaussian distribution (GGD) random variables. We will show in Section III that the potential coding gain of vector quantization heavily depends on bit rate and the parameters of GGD, or more general, on "tails" of distribution. The well- known estimate of vector quantization gain of 1.54 dB (for the MSE as a distortion criterion) is valid under the assumption of high rates or small distortions (see [9], [10]). It appears that for distributions with heavy tails the distortions become "small" for much higher rates than for, say, Gaussian distribution. We will show in Section III that for GGD with small value of the parameter $\alpha$ for coding rates below 1 bit per sample the rate-distortion function of the scalar quantization is very close to that of the vector quantization, i.e. to the theoretical limit.

This phenomenon makes the potential gain of vector quantization rather small and vector quantization loses against scalar quantization in terms of distortion/complexity tradeoff.

We would consider closeness to theoretical limits on quantization performance as a figure of merit of concrete quantization scheme. For stationary random process such theoretical limits are defined by Shannon rate-distortion function. There are several obstacles when using this approach in real applications. They are: non-stationarity of real signals, complexity of their mathematical models, and the absence of analytical expressions for rate-distortion functions for most probability density functions.

Our approach to surmount these problems is similar to that used in universal lossless source coding [13]. We split spectral coefficients into subbands, assuming that the spectrum coefficients are stationary in subbands. For each subband, we estimate the parameters of GGD model, as it is illustrated in Section II. Parameters are estimated based on moment's method. Then, we construct a quantizer for the source with GGD using these estimates instead of the unknown true values of parameters. In the theory of universal lossless source coding it is proven that similar strategy provides the rate arbitrarily close to the entropy of the source [13]. The redundancy of universal coding per encoded letter is proportional to $(\log_2 n)/n$ and vanishes with the increasing length $n$ of the sequence used for estimating the unknown parameters. This redundancy value is interpreted as the cost of side information. The examples of side information are scale index, scale factor, etc that can strongly depend on type of quantizer.

For the GGD sources we show that a near optimum scalar quantizer can be found among uniform quantizers or extended zero-zone (EZZ) quantizers (the efficiency of optimum scalar quantization and uniform scalar quantization for GGD variables was studied in [6]). This means that not much side information about the quantizer needs to be transmitted for a given subband of a frame of the encoded data: only the quantization step and the relative width of the zero zone. Thereby, like in universal data compression, the cost of side information is relatively small, even for rather short quantized sequences.

The rest of the paper is organized as follows. In Section II the source model is considered. In Section III we consider types of scalar quantizers and study their performance. Section IV is devoted to EZZ quantizers. Adaptation of quantizer parameters to changing input data distribution is studied in Section V.

## II. SOURCE MODEL

We consider the sequence of spectrum coefficients of each separate spectrum subband as a stationary sequence of independent identically distributed random variables. Thus the source model is fully described by a one-dimensional probability density function.

In multimedia applications like video- and audio- data coding, the generalized Gaussian distribution (GGD) is often used as a source model. The corresponding probability density function has the form

$$f(x) = \frac{\alpha \eta(\alpha,\sigma)}{2\Gamma(1/\alpha)} \exp\left\{-\left(\eta(\alpha,\beta)|x-m|\right)^\alpha\right\},$$

where $m$, $\sigma$ are mathematical expectation and standard deviation of the random variable, $\alpha$ is the parameter, $\Gamma(\cdot)$ is the Gamma function

$$\Gamma(x) = \int_0^\infty t^{x-1} e^{-t} dt, \ x > 0,$$

and

$$\eta(\alpha,\sigma) = \sigma^{-1}\left[\frac{\Gamma(3/\alpha)}{\Gamma(1/\alpha)}\right]^{1/2}.$$

Plots of $f(x)$ are shown in Fig.2. Special cases of GGD are Gaussian distribution ($\alpha = 2$), Laplacian distribution ($\alpha = 1$) and uniform distribution ($\alpha \to \infty$).

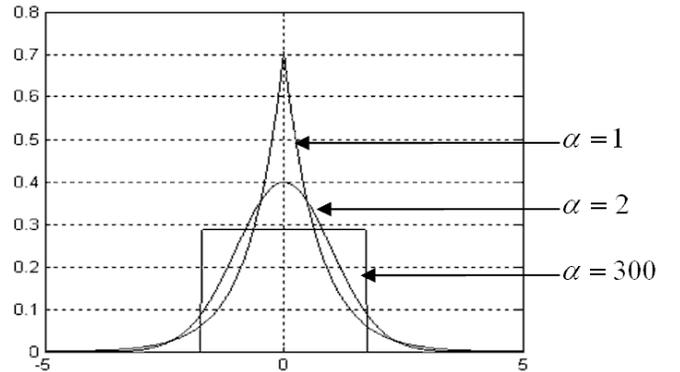

Fig. 2. Generalized Gaussian distribution.

The information-theoretical rate-distortion function $R(D)$ [7] for a memoryless discrete-time stationary random process is defined as

$$R(D) = \min_{f(y|x): E[d(x,y)] \le D} I(X;Y),$$

where $X$ and $Y$ are the source alphabet and the approximation alphabet, respectively,

$$I(X;Y) = \iint_{X\ Y} f(x) f(y|x) \log_2 \frac{f(y|x)}{f(x)} dxdy$$

is mean mutual information between $X$ and $Y$, and $d(x,y)$ is a nonnegative function which is called distortion measure. We consider the mean squared error (MSE)

$$d(x,y) = (x-y)^2$$

as a fidelity criterion.

The rate-distortion function $R(D)$ determines the least achievable bit rate $R = R(D)$ under restriction that the mean distortion measure does not exceed $D$. We exploit numerical method [8] for computing $R(D)$.

To verify whether the GGD-model is appropriate for audio coding, we have done the following experiments. We split the MDCT spectrum coefficients of audio signals into subbands according to Bark scale and for each subband we generated a long data sequence obtained by applying an MDCT-based filterbank to a large bank of audio fragments. Then for each subband $x_1,...,x_n$ the parameters of GGD were estimated. We assumed average value $m = 0$, and estimated variance as

$$\hat{\sigma}^2 = \frac{1}{n}\sum_{i=1}^{n} x_i^2,$$

and the first absolute moment as

$$\hat{\mu} = E[|x|] = \frac{1}{n}\sum_{i=1}^{n}|x_i|.$$

The estimated parameter $\hat{\alpha}$ of GGD is computed as a solution of the equation

$$\frac{\hat{\sigma}^2}{\hat{\mu}^2} = \frac{\Gamma(1/\alpha)\Gamma(3/\alpha)}{\Gamma^2(2/\alpha)}. \quad (1)$$

The Blahut algorithm [8] was used to compute two rate-distortion functions for each subband:
1) "Theoretical" function $R_T(D)$ was computed for discrete alphabet $\hat{X}$ obtained by fine quantization of $X$ with probabilities of $x \in \hat{X}$ found using GGD with $\alpha$ determined as a solution of (1).
2) "Empirical" function $R_E(D)$ was computed for discrete alphabet $\hat{X}$ obtained by fine quantization of $X$ with probabilities of $x \in \hat{X}$ found as estimated probabilities directly from the real sample data sequence.

One typical example of these two functions is given in Fig. 3 for estimated GGD parameters $\alpha = 0.67$ and $\sigma^2 = 1$. It is clear from the figure that the two curves are almost indistinguishable, which confirms that GGD is a good mathematical model for MDCT spectrum coefficients.

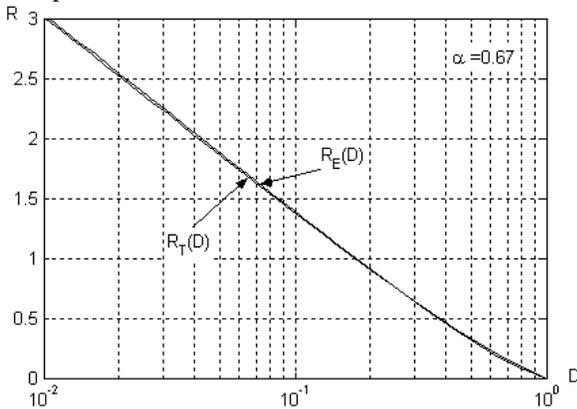

Fig. 3. Theoretical and empirical rate-distortion functions.

### III. UNIFORM, OPTIMAL UNIFORM AND NON-UNIFORM SCALAR QUANTIZER

We will start with *upper bounds* on the rate-distortion function for scalar quantization.

Most of the estimates of quantization performance are derived for so-called "high resolution" quantization. Under the assumption that the number of quantization levels is so large that the probability density function is almost uniform at each quantization step, the following estimate of the scalar quantization rate-distortion function $R_S(D)$ was obtained by Koshelev [9] and later by Gish and Pierce [10]:

$$R_S(D) \leq R_{SH}(D) + \frac{1}{2}\log_2\frac{\pi e}{6} \approx R_{SH}(D) + 0.2546\, bits, \quad (2)$$

where $R_{SH}(D)$ is the lower Shannon bound on rate-distortion function, which can be written in the form

$$R(D) \geq H_0(X) - \frac{1}{2}\log_2(2\pi e D) = R_{SH}(D). \quad (3)$$

In the above formula $H_0(X)$ is the differential entropy, which for GGD equals

$$H_0(X) = -\log_2\left[\frac{\alpha\eta(\alpha,\sigma)}{2\Gamma(1/\alpha)}\right] + \frac{1}{\alpha\ln 2}.$$

Achievable scalar quantization performance was thoroughly investigated by Farvardin and Modestino in [6], using Lagrangian minimization of the MSE, under restriction on the bit rate (more exactly, on the entropy of the approximation alphabet) over all quantizer parameters, for different number of quantization levels. It follows from [6] that optimum uniform quantization performances almost coincide with optimum quantization performance for broad class of probability distributions. This is not surprising, since it was established analytically by T. Berger [11] that for GGD with $\alpha=1$ (Laplacian distribution) optimum uniform quantization is entropy-optimal.

Rate-distortion functions for four generalized Gaussian distributions are shown in Figs. 4–7 together with performance of scalar quantizers. In each figure, $R(D)$ denotes the rate-distortion functions obtained using Blahut algorithm; $R_{SH}(D)$ denotes Shannon bound (3), and the dotted line shows the Koshelev bound (2). Curves denoted by $R_{USQ}(D)$ and $R_{OUSQ}(D)$ represent performance of uniform scalar quantization (USQ) and optimal uniform scalar quantization (OUSQ), respectively. Here, USQ implies that quantization procedure is performed as division of input data by the fixed quantization step (properly chosen to provide required bit rate) followed by rounding. Middle points of quantization intervals are used as the reconstruction values.

The OUSQ differs from USQ in the reconstruction values which are computed as mass centers of quanta. We do not show $R_S(D)$ since it is indistinguishable from $R_{OUSQ}(D)$.

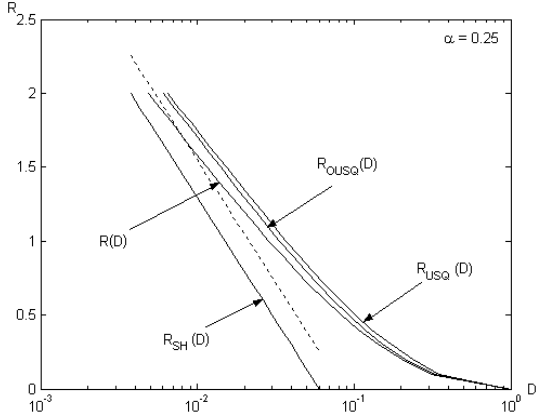
Fig. 4. Rate-distortion function for $\alpha = 0.25$.

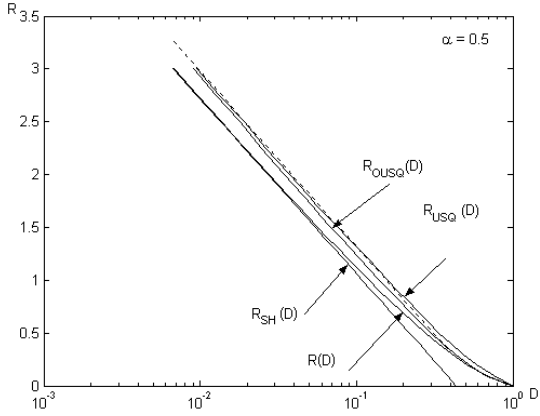
Fig. 5. Rate-distortion function for $\alpha = 0.5$.

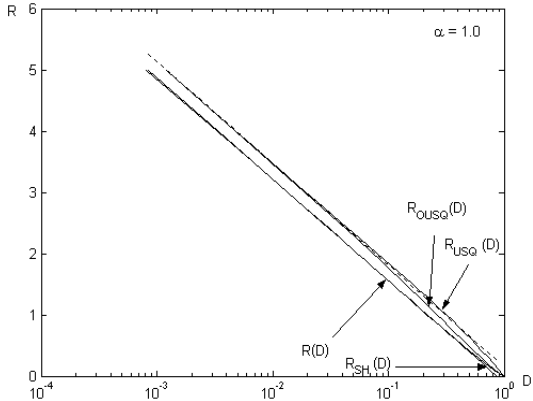
Fig. 6. Rate-distortion function for $\alpha = 1.0$.

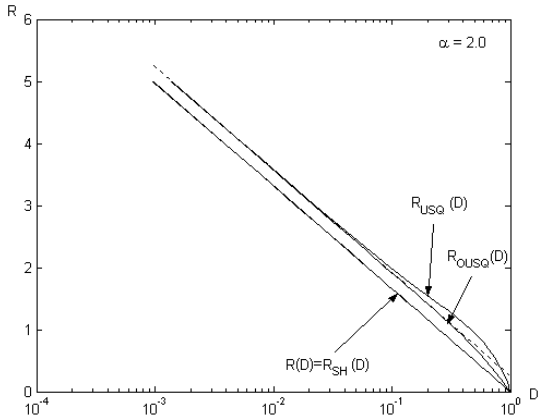
Fig. 7. Rate-distortion function for $\alpha = 2.0$.

It is easy to see from the plots that the behavior of the rate-distortion functions for small α differs significantly from their behavior for Laplacian (*α*=1) and Gaussian distribution (*α*=2). In particular, for bit rates below 1 bit per sample scalar quantization is closer to theoretical minimum *R*(*D*) for small α than for large α. For example, the distortion level of $10^{-2}$ (20 dB) for Gaussian distribution ($\alpha = 2$) can be achieved at bit rate of approximately 3.33 bits/sample using vector quantization, or at rate of 3.58 bits/sample with optimal scalar quantization. The same distortion level for GGD with $\alpha = 0.25$ can be achieved at 1.50 bits/sample using vector quantization and at rate of 1.61 bits/sample using scalar quantization. Therefore, theoretically achievable gain of vector quantization 0.25 bits/sample cannot be achieved for typical sequences of transform coefficients of audio signals.

Although uniform quantization itself can be easily implemented, the reconstruction is not so simple, because it requires storing reconstruction values for all quantization intervals. It could be possible to keep a full set of reconstruction levels if the bit rate is fixed and input signal is a stationary process. However, neither of the two conditions is valid in audio coding, and thus, using optimal uniform quantization became too complicated. We have found a much simpler solution that yields near-optimal scalar quantization.

## IV. EZZ SCALAR QUANTIZER

In this section we present simulation results for quantizer with the extended zero zone (EZZ). The set of quantization thresholds can be described by the following set of numbers:
$$B(j,\lambda)=\{\pm \lambda 2^{j-1}, \pm \lambda(2^{j-1}+1), \pm \lambda(2^{j-1}+2),\ldots\}, \quad (4)$$
where $\lambda>0$ is the scaling factor and *j* is the parameter which determines the size of the zero zone, $j \in \{0,1,\ldots\}$.

It is clear from (4) that all thresholds are equally spaced with interval $\lambda$ except the two thresholds $\pm\lambda 2^{j-1}$ whose distance is $\lambda 2^j$. This value is the size of the zero zone. Obviously, the scale $B(0,\lambda)$ corresponds to the uniform quantization. Examples of the scales $B(j,\lambda = 1)$ are shown in Fig. 8.

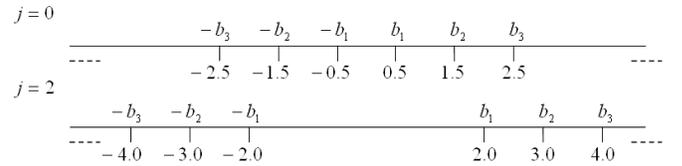

Fig. 8. Examples of quantization scales.

Now let us choose the set of approximating values. We consider the following sets of quantizers:
1) EZZ, with scales defined by (4) and approximating values placed in the middle of each quantization interval;
2) OEZZ (Optimized EZZ) with scales defined by (4) and optimal approximating values placed into gravity mass center of the probability density function of each quantization interval;
3) SOEZZ (Sub-optimal EZZ) with scales defined by (4) and

optimal approximating values for 2 intervals closest to the zero interval and all other approximating values placed in the middle points of the corresponding intervals.

For a fixed bit rate $R$ the corresponding distortion levels are related as

$$D_{EZZ}(R) \geq D_{SOEZZ}(R) \geq D_{OEZZ}(R) \qquad (5)$$

while the complexities and the amount of side information required for describing these quantizers are related in the opposite manner. Our goal is to estimate the gap between distortion values in (5) for the GGD random variables.

The *quantization gain* is defined as

$$G = 10\log_{10}\frac{\sigma^2}{D} \, (dB), \qquad (6)$$

where $\sigma^2$ is the source data variance and $D$ is quantization error variance.

In (6) without loss of generality we can set $\sigma^2 = 1$. Then the maximum achievable gain $G_{max}(R)$ under a fixed bit rate $R$ can be computed as

$$G_{max}(R) = -10\log_{10} D_0,$$

where $D_0$ is the solution of the equation

$$R(D_0) = R$$

and $R(D)$ is the rate-distortion function.

If

$$G(R) = -10\log_{10} D(R)$$

denotes the quantization gain of some quantizer then we call the difference

$$L(R) = G_{max}(R) - G(R)$$

the "*loss of coding gain with respect to the theoretical limit*".

Plots of $L(R)$ for different quantizers and for different values of parameter $\alpha$ are shown in Fig 9. For $\alpha = 1$ and $\alpha = 2$ these plots are obtained numerically and for smaller $\alpha$ they are obtained by simulation.

It follows from these plots that SOEZZ is very close to optimal scalar quantization for all distributions and especially for small $\alpha$ which are typical for spectrum quantization problem. Note also that the gain of SOEZZ with respect to uniform quantization is rather high. In particular, for small $\alpha$ this gain approaches 0.5 dB, while for large $\alpha$ the gain achieves 1 dB.

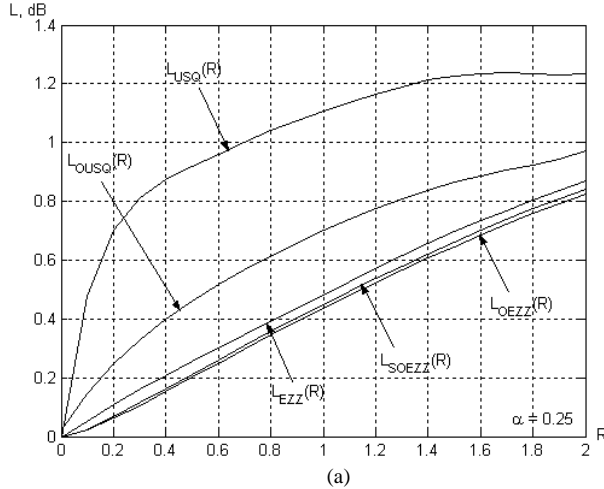

(a)

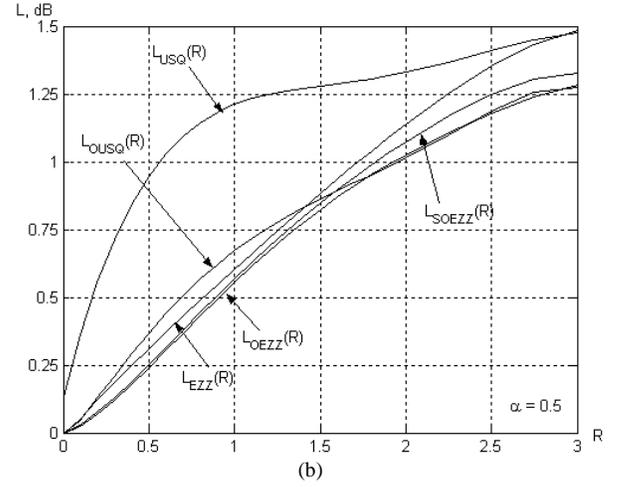

(b)

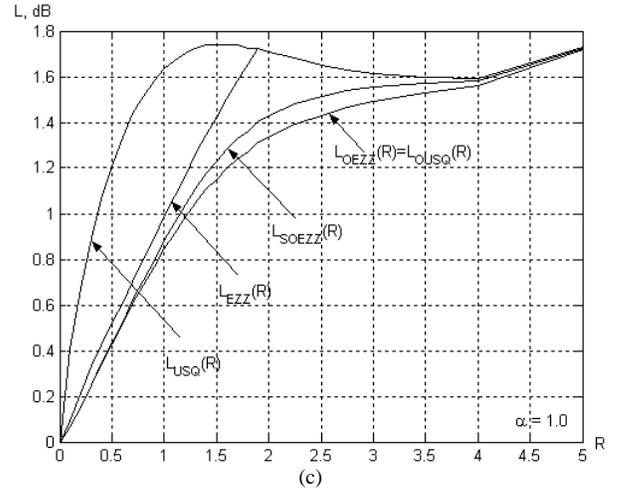

(c)

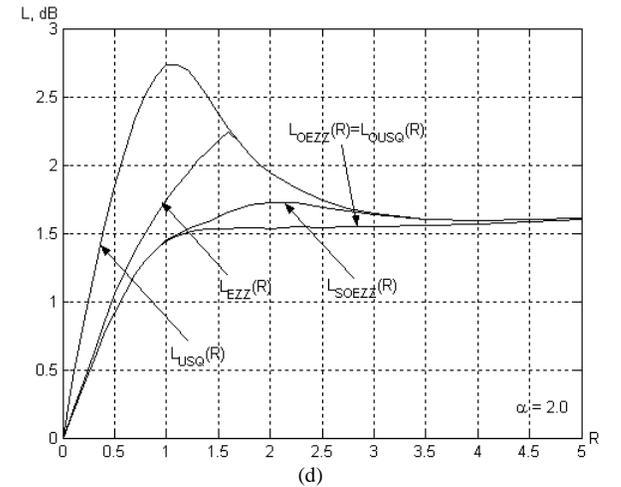

(d)

Fig. 9. Loss of coding gain with respect to theoretical limit. (a) GGD distribution with parameter $\alpha = 0.25$, (b) GGD distribution with parameter $\alpha=0.5$, (c) GGD distribution with parameter $\alpha = 1.0$, (d) GGD distribution with parameter $\alpha = 2.0$

Notice again that the gain loss with respect to vector quantization is near 0.5 dB only for bit rates about 1 bit/sample. The efficiency of known vector quantization and trellis quantization schemes with reasonable complexities is also roughly 0.5 dB below the theoretical limit.

## V. Adaptive scalar quantization

It follows from the above considerations that near-optimal quantization performance can be achieved using SOEZZ which, according to (4), can be completely described by 3 parameters: quantization step $\lambda$, zero zone width $j$, and approximating value $a_1$ for the first non-zero quantization interval. The EZZ quantizer has slightly worse results than SOEZZ and can be described by two parameters: quantization step $\lambda$, and zero zone width $j$. Both EZZ and SOEZZ quantizer can be efficiently used for practical implementation. The OEZZ quantizer has best results, but it is not used since the number of of parameters to be transmitted to decoder as side information is large (as many as number of quanta in a scale).

The following approach can be used for estimating the EZZ, SOEZZ or OEZZ quantizer parameters.

The rate-distortion functions for typical values of $\alpha$ for the EZZ, SOEZZ or OEZZ quantizer have to be known to the encoder. These functions can be kept in the form of data arrays or as simple approximate analytical expressions (e.g. interpolation polynomials). A typical function of SOEZZ is shown in Fig. 10 for $\alpha = 0.5$, $\sigma = 1$. These functions for EZZ and OEZZ can differ only in regions of scale index using. Notice that for each point $(R,D)$ of the rate-distortion function the optimal parameters $(\lambda,j)$ are known to the encoder.

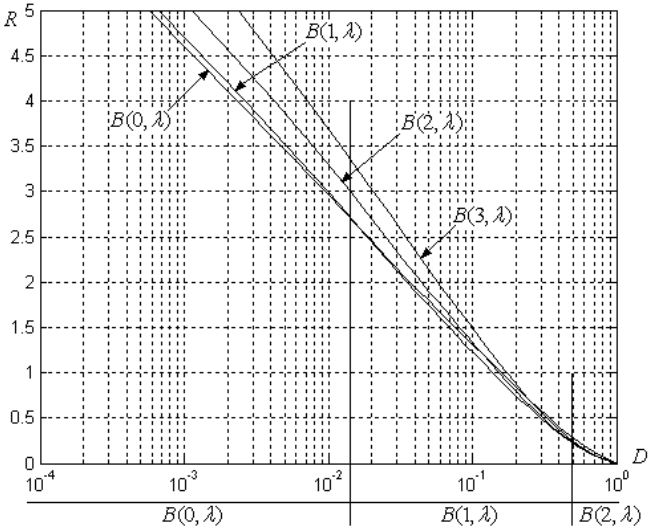

Fig. 10. Parameters of the SOEZZ quantizer.

Let $(x_1,...,x_n)$ be the data sequence to be quantized. The adaptive quantization procedure for EZZ quantizer is shown in Fig.11, and can be modified to use with any type of extended zero-zone quantizer. For SOEZZ, we need additional estimate reconstruction value of first non-zero quantum. For OEZZ quantizer, we need estimate reconstruction value of all non-zero quanta.

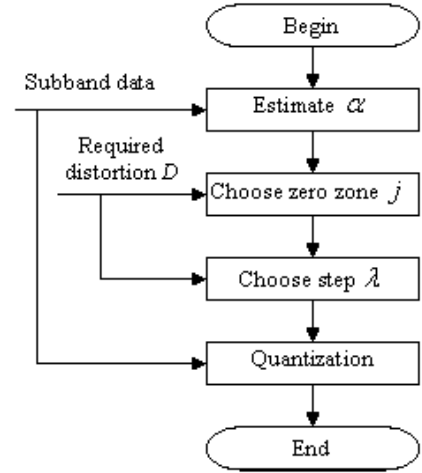

Fig. 11. Adaptive quantization.

Procedure starts with estimating GGD parameters for which we use the same approach as in [12]. First, the variance

$$\hat{\sigma}^2 = \frac{1}{n}\sum_{i=1}^{n} x_i^2$$

and the first absolute moment

$$\hat{\mu} = E[|x|] = \frac{1}{n}\sum_{i=1}^{n}|x_i|$$

estimates are computed. The estimated parameter $\hat{\alpha}$ of GGD is found as a solution of the equation

$$\frac{\hat{\sigma}^2}{\hat{\mu}^2} = \frac{\Gamma(1/\alpha)\Gamma(3/\alpha)}{\Gamma^2(2/\alpha)}.$$

To compute the quantizer parameters the required distortion level $D$ obtained from PAM module must be normalized by $\hat{\sigma}^2$, and the function $R(D)$ corresponding to the estimated $\hat{\alpha}$ has to be used for evaluating $j$ and $\lambda$.

If SOEZZ quantizer is considered then one more calculation is needed. After quantization, the reconstruction level $a_1$ can be computed as follows

$$a_1 = \frac{1}{n_1}\sum_{i:|x_i|\in[b_1,b_2]}|x_i|,$$

where the sum is computed over all $i$ such that $|x_i|$ belong to the first non-zero quantization interval and $n_1$ is the number of elements in this sum.

If OEZZ quantizer is considered then following calculations are required. After quantization, the reconstruction level of each non-zero quantum $a_j$ can be computed as follows

$$a_j = \frac{1}{n_j}\sum_{i:|x_i|\in[b_j,b_{j+1}]}|x_i|$$

where the sum is computed over all $i$ such that $|x_i|$ belong to the $j$-th non-zero quantization interval and $n_j$ is the number of elements in this sum.

## VI. Conclusions

We analyzed efficiency of scalar quantization followed by entropy coding, used for encoding filterbank outputs of an audio codecs. The GGD is used as a model of one-dimensional probability distribution model of the data to be quantized. Using the Blahut algorithm for evaluating the rate-distortion function we have shown that the theoretically achievable efficiency computed from the model virtually coincides with the empirical rate-distortion function obtained directly from the long audio data sequence. Thereby we justify the choice of the GGD as a source model.

The potential efficiency of scalar quantization was estimated and compared with vector quantization efficiency. For typical audio data the gain of vector quantization over scalar quantization is rather small. Moreover, the efficiency of the uniform scalar quantization and the extended zero-zone (EZZ) quantization is close to that of optimum scalar quantization.

The important advantage of uniform and EZZ quantization is that they can be described by a small number of parameters. Therefore EZZ quantization used for adaptive quantization does not require transmitting a large amount of side information.

## REFERENCES

[1] J.P. Princen and A.B. Bradley, "Analysis/synthesis filterbank design based on time domain aliasing cancellation," *IEEE Trans. Acoust., Speech Signal Processing,* vol. ASSP-34, pp. 1153-1161, Oct., 1986.
[2] H. S. Malvar, *Signal processing with lapped transforms.* Artech House., 1992.
[3] M. Temerinec and B. Edler, "LINC: A common theory of transform and subband coding," *IEEE Trans. Commun.,* vol. COM-41, 2, pp. 266-274, Feb. 1993.
[4] R.M. Gray and D.L. Neuhoff, "Quantization," *IEEE Trans. Inform. Theory*, vol. IT-44, No 6, pp. 2325 –2383, Oct. 1998.
[5] T. Berger and J. Gibson, "Lossy source coding," *IEEE Trans. Inform. Theory*, v. IT-44, No 6, pp. 2702-2703, Oct, 1998.
[6] N. Farvardin and J. W. Modestino, "Optimum quantizer performance for a class of non- Gaussian memoryless Sources," *IEEE Trans. Inform. Theory*, v. IT-30, No 3, pp. 485- 497, May, 1984.
[7] J. A. Thomas and T. M. Cover, *Elements of Information Theory,* John Wiley&Sons, New York, 1981.
[8] R. E. Blahut, "Computation of Channel Capacity and Rate-Distortion Functions", *IEEE Trans. Inform. Theory*, IT-18, No 4, pp. 460-473, Jul., 1972.
[9] V. N. Koshelev, "Quantization with minimum entropy," *Probl. Inform. Transmiss.* v.14, pp. 151-156, 1963 (Section X).
[10] H. Gish and J.N. Pierce, "Asymptotically efficient quantizing," *IEEE Trans. Inform. Theory,* v.IT-14, No 5, pp. 676-683, Sept., 1968.
[11] T. Berger, "Optimum quantizers and permutation codes," *IEEE Trans. Inform. Theory,* v. IT-18, No 6, pp. 759-765, Nov. 1972.
[12] K. Sharifi K. and A. Leon-Garcia, "Estimation of shape parameter for generalized Gaussian distributions in subband decompositions of video," *IEEE Trans. on Circuits and Systems for Video Technology*, 5(1), pp. 52-56, Febr. 1995.
[13] L. D. Davisson, "Universal noiseless coding," *IEEE Trans. Inform. Theory,* v. IT-19, No 6, pp. 783-795, Nov. 1973.

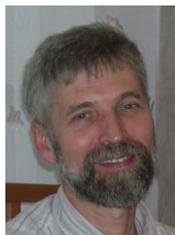
**Boris D. Kudryashov** was born in Leningrad, USSR (now St.-Petersburg, Russia) on July, 9, 1952. He received the Diploma degree in electrical engineering in 1974, the Ph.D. degree in technical sciences degree in 1978 both from the Leningrad Institute of Aerospace Instrumentation (LIAP) and the Doctor of Science degree from the Institute of Information Transmission Problems (IPPI), Moscow in 2005.

Since 1978 he was been first Assistant Professor and then Associate Professor and Professor at the State University of Aerospace Instrumentation (former LIAP), St.-Petersburg, Russia. Since 2007 he is Professor of St.-Petersburg State University of Information Technology, Mechanics and Optics, St.-Petersburg, Russia.

His research interests include coding theory, information theory and applications to speech, audio, and image coding. He has published more than 70 papers in journals and proceedings of international conferences, 15 US patents and published patent applications in image, speech and audio coding.

Prof. Kudryashov served as a member of Organizing Committees of ACCT International Workshops.

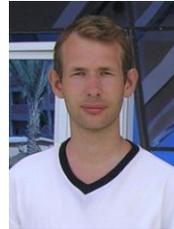
**Anton V. Porov** was born in Leningrad, USSR (now St.-Petersburg, Russia) on January, 12, 1980. He received the Diploma degree in Computer science 2003.

Since 2003 he was been an Engineer at State University of Aerospace Instrumentation (GUAP), St.-Petersburg, Russia. In 2005-2007 he was member of Assistant Staff of Samsung Advanced Institute of Technology (SAIT), Suwon-si, South Korea. Since 2008 he is a research engineer at St.-Petersburg State University of Information Technology, Mechanics and Optics, St.-Petersburg, Russia. His research interests include coding theory, information theory and application to speech and audio coding.

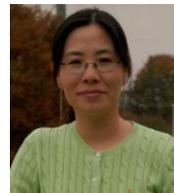
**Eunmi L. Oh** received the Ph.D degree in psychology from the University of Wisconsin – Madison, WI, USA in 1997. Her major was psychoacoustics focusing on masking models. She is currently working at Samsung Advanced Institute of Technology, Yongin, South Korea. Her recent work is concerned with perceptual coding and scalable audio coding.